\documentclass[graybox]{svmult}

\usepackage{mathptmx}       
\usepackage{helvet}         
\usepackage{courier}        
\usepackage{type1cm}        
\usepackage{makeidx}         
\usepackage{graphicx}        
\usepackage{multicol}        
\usepackage[bottom]{footmisc}

\makeindex             

\begin{document}

\title*{Cell death and  life in cancer: mathematical modeling of cell fate decisions}
\author{Andrei Zinovyev, Simon Fourquet, Laurent Tournier, Laurence Calzone and Emmanuel Barillot}
\institute{Andrei Zinovyev \at U900 INSERM/Institut Curie/Ecole de
Mines, Institut Curie, 26 rue d'Ulm, Paris, France, 75005
\email{andrei.zinovyev@curie.fr}
\and Simon Fourquet \at
U900 INSERM/Institut Curie/Ecole de Mines,
Institut Curie, 26 rue d'Ulm, Paris, France, 75005
\email{simon.fourquet@curie.fr}
\and Laurent Tournier \at INRA, Unit MIG (Math\'ematiques, Informatique et G\'enome),
Domaine Vilvert, Jouy en Josas, France, 78350,
\email{laurent.tournier@jouy.inra.fr}
\and Laurence Calzone \at
U900 INSERM/Institut Curie/Ecole de Mines, Institut Curie, 26 rue
d'Ulm, Paris, France, 75005 \email{laurence.calzone@curie.fr}
\and Emmanuel Barillot \at
U900 INSERM/Institut Curie/Ecole de Mines, Institut Curie, 26 rue
d'Ulm, Paris, France, 75005 \email{emmanuel.barillot@curie.fr}}
%
%
\maketitle

\abstract*{Tumor development is characterized by a compromised
balance between cell life and death decision mechanisms, which are
tightly regulated processes in normal cells. Understanding this
process provides insights for developing new treatments for
fighting with cancer. We present a study of a mathematical model
describing cellular choice between survival and two alternative
cell death modalities: apoptosis and necrosis. The model is
implemented in discrete modeling formalism and allows to predict
probabilities of having a particular cellular phenotype in
response to engagement of cell death receptors. Using an original
parameter sensitivity analysis developed for discrete dynamic
systems, we determine
variables that appear to be critical in the cellular fate decision
and discuss how they are exploited by existing cancer therapies.}

\abstract{Tumor development is characterized by a compromised
balance between cell life and death decision mechanisms, which are
tighly regulated in normal cells. Understanding this process
provides insights for developing new treatments for fighting with
cancer. We present a study of a mathematical model describing
cellular choice between survival and two alternative cell death
modalities: apoptosis and necrosis. The model is implemented in
discrete modeling formalism and allows to predict probabilities of
having a particular cellular phenotype in response to engagement
of cell death receptors. Using an original parameter sensitivity
analysis developed for discrete dynamic systems, we determine
variables that appear to be critical in the cellular fate decision
and discuss how they are exploited by existing cancer therapies.}

\section{Introduction}

Evading various programmed cell death modalities is considered as
one of the major hallmarks of cancer cells \cite{Weinberg2011}. A
better understanding of the pro-death or prosurvival roles of the
genes associated with various cancers, and their interactions with
other pathways would set a ground for re-establishing a lost death
phenotype and
identifying potential drug targets.

Recent progress in studying the mechanisms of cell life/death
decisions revealed its astounding complexity. Among many, one can
mention three difficulties on the way to characterize, describe
and create strict mathematical descriptions of these mechanisms.

First, the signaling network allowing a cell to react to an
external stress (such as damage of DNA, nutrient and oxygen
deprivation, toxic environment) is assembled from highly redundant
pathways which are able to compensate each other in one way or
another. For example, there exist at least seven distinct and
parallel survival pathways associated with action of AKT protein
\cite{McCormick04}. Disruption of one of these pathways in a
potential cell death-inducing cancer therapy can be in principle
compensated by the others. Thus, understanding and modeling the
survival response in its full complexity is a daunting task.

Second, cellular death is an extremely complex phenotype
that cannot merely be described as a simple disaggregation of
cellular components driven by purely thermodynamical laws. Several
distinct modes of cell death were identified in the last
decade \cite{Kroemer2008}, such as necrosis, apoptosis and
autophagy. Importantly, all these cell death modalities are
controlled by cellular biochemical mechanisms, activated in
response to diverse types of stress: roughly speaking, a cell is
usually pre-programmed to die in a certain manner, sending
appropriate signals to its surroundings so as to limit tissue
toxicity and allow recycling of its components. {\it Necrosis} is
a type of cell death usually associated with a lack of important
cellular resource such as ATP, which makes functioning of many
biochemical pathways impossible. This is why it was long thought
of as an uncontrolled and purely thermodynamics-driven degradation
of cellular structures. However, recent research showed that
necrosis can be triggered by specific signals through the
activation of tightly regulated pathways, and can even proceed
without ATP depletion \cite{Kroemer2008}. By contrast, {\it
apoptosis} as a form of cellular suicide was, from the very
beginning, described as a mode of cell death requiring energy for
the permeabilization of mitochondrial membranes and cleavage of
intracellular structures. {\it Autophagy} remains a relatively
poorly understood cell death mechanism, which seems to serve both
as a survival or a death modality. Upon certain stress conditions,
and until this stress is relieved, cellular components such as
damaged proteins or organelles are digested and recycled into
reusable metabolites, and metabolism is reoriented so as to spare
vital functions. Long lasting, non-relievable stress was described
as triggering autophagic cell death, through unaffordable cellular
self-digestion. However, no experimental evidence ever
unambiguously demonstrated that such cell death is directly
executed by autophagy in vivo, but in the special case of the
involution of {\it Drosophila melanogaster} salivary glands
\cite{Kroemer2008}.

The third difficulty can be attributed not directly to the
complexity of the biochemical mechanisms but rather to our
capabilities of apprehending the design principles used by
biological evolution. Inspired by engineering practices, we tend
to investigate complex systems by splitting them into relatively
independent modules and associating well-characterized
non-overlapping functions to each molecular detail. Applying such
reductionist approaches to biology comes with a caveat. Most
cellular molecular machineries cannot be naturally dissected or
associated with well-defined functions, and sets of overlapping
functions can be distributed among groups of molecular players.

Not having the ambition to deal with the whole complexity of cell
fate decisions {\it in vivo}, we decided to concentrate on
modeling the outcome of a classical and rather well-defined
experiment of inducing cell death: adding to a cell culture
specific ligands (Tumor Necrosis Factor, TNF, or other members of
its family such as FASL). These so-called death ligands can engage
death receptors and trigger apoptosis or necrosis, or activate
pro-survival mechanisms \cite{Herreweghe2010}. The net outcome of
such experiments depends on many circumstances: cell type, dose of
the ligand, duration of the treatment, specific mutations in cell
genomes, etc. Moreover, it is believed that the outcome can have
intrinsic stochastic nature governed by cellular decision making
mechanisms and intrinsic molecular noise \cite{Balazsi11}. Trying
to characterize the biochemical response of a cell to this
relatively simple kind of perturbation allows to understand
certain cell fate decision mechanisms.

In this paper, we briefly describe and carefully analyze a
mathematical model of cell fate decision between
survival and two alternative modes of cell death:
apoptosis and necrosis. The model was created and introduced in
\cite{Calzone10}. Here propose the principles for wiring and
parametrizing a biological diagram that describes this cellular
switch. In addition to \cite{Calzone10}, here, by applying a novel
sensitivity analysis specifically developed for discrete modeling,
we identify fragile sites of the cell fate decision mechanism. In
conclusion, we compare our analysis with our current knowledge of
cellular decision making fragilities utilized by cancer and cancer
therapies.

\section{Mathematical model of cell fate decision}

\begin{figure}[htbp]
\center{\includegraphics[width=100mm]{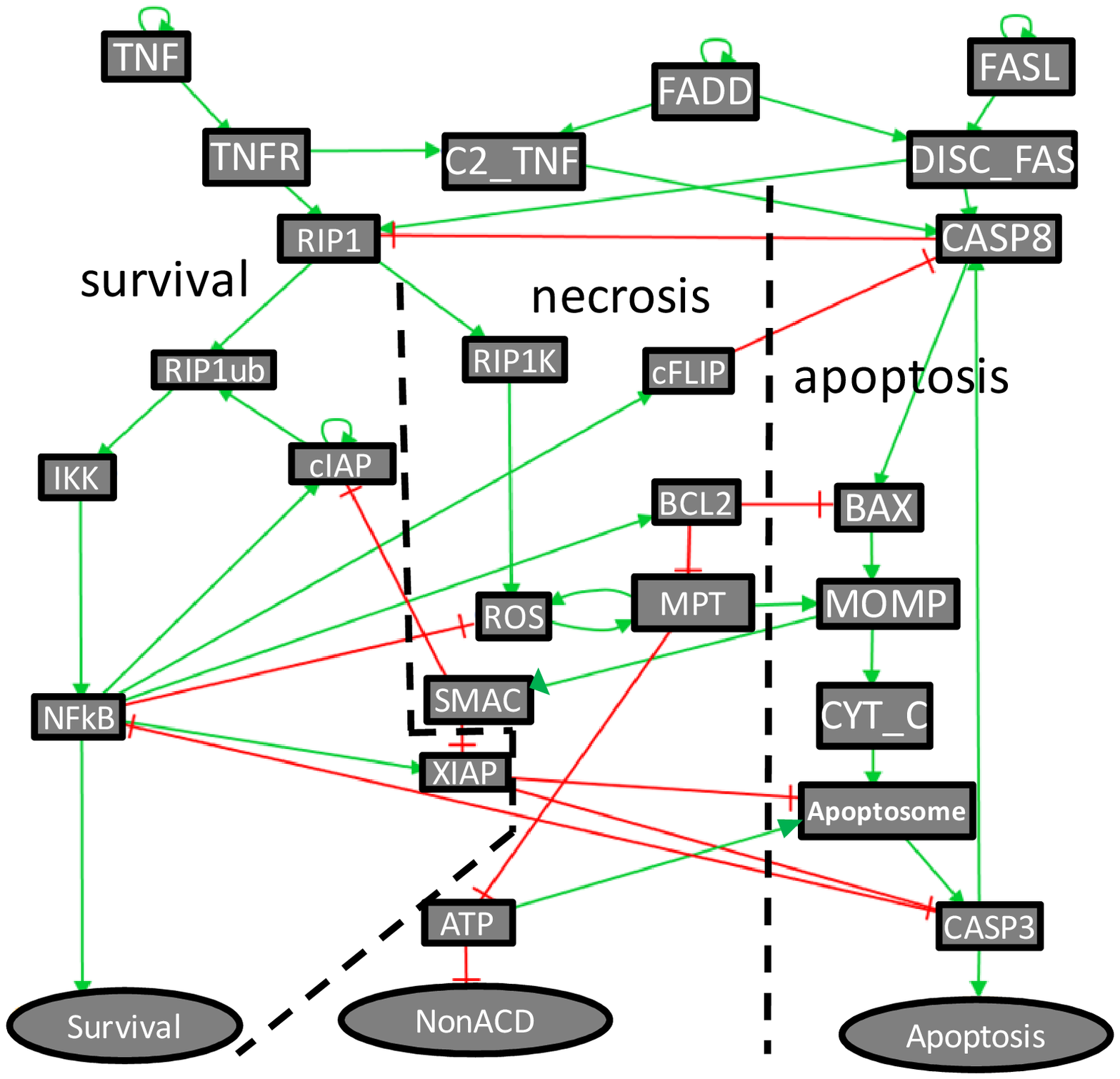}
\vspace{1cm}\scriptsize{
\begin{tabular}{|l|l}
DISC\_TNF' = TNFR {\bf \tiny AND} FADD & TNF' = TNF \\ RIP1' =
(TNFR {\bf \tiny OR} DISC\_FAS) {\bf \tiny AND} ({\bf \tiny NOT}
CASP8) & FADD' = FADD \\ CASP8' = (DISC\_TNF {\bf \tiny OR}
DISC\_FAS {\bf \tiny OR} CASP3) {\bf \tiny AND} ({\bf \tiny NOT}
cFlip) & FAS' = FAS \\ RIPub' = RIP1 {\bf \tiny AND} cIAP & TNFR'
= TNF \\ cIAP' = (NFkB {\bf \tiny OR} cIAP) {\bf \tiny AND} ({\bf
\tiny NOT} SMAC) & RIP1K' = RIP1 \\ BAX' = CASP8 {\bf \tiny AND}
({\bf \tiny NOT} BCL2) & cFlip' = NFkB \\ ROS' = (RIP1K {\bf \tiny
OR} MPT) {\bf \tiny AND} ({\bf \tiny NOT} NFkB) & IKK' = RIP1ub \\
MPT' = ROS {\bf \tiny AND} ({\bf \tiny NOT} BCL2) & BCL2' = NFkB
\\ MOMP' = BAX {\bf \tiny OR} MPT & SMAC' = MOMP
\\ NFkB' = IKK {\bf \tiny AND} ({\bf \tiny NOT} CASP3) & CYT\_C' = MOMP
\\ XIAP' = NFkB {\bf \tiny AND} ({\bf \tiny NOT} SMAC) & DISC\_FAS' = FAS {\bf \tiny AND} FADD \\ Apoptosome' = CYT\_C {\bf \tiny AND} ATP {\bf \tiny AND}
({\bf \tiny NOT} XIAP) & ATP' = {\bf \tiny NOT} MPT \\ CASP3' =
Apoptosome {\bf \tiny AND} ({\bf \tiny NOT} XIAP) & \\
\end{tabular}}
\caption{ \label{Model} Biological diagram of molecular
interactions involved in cell fate decisions derived from the
biological literature. The diagram is roughly divided by dashed
lines into three modules corresponding to three submechanisms of
cell fate decisions. Notations: 1) Proteins: TNF, FADD, FASL,
TNFR, CASP8, cIAP, cFLIP, BCL2, BAX, IKK, NF$\kappa$B, CYT\_C,
SMAC, XIAP, CASP3; 2) States of proteins: RIP1ub
(ubiquitinated form of RIP1), RIP1K (kinase function of RIP1); 3)
Small molecules: ATP, ROS (Reactive oxygen species); 4) Molecular
complexes: Apoptosome, C2\_TNF, DISC\_FAS; 5) Molecular processes:
MPT (Mitochondria permeabilization transition), MOMP
(Mitochondrial outer membrane permeabilization); 6) Phenotypes:
Survival, Apoptosis, NonACD (Non-apoptotic cell death). Below the
table of logical rules defining the discrete mathematical model is
provided.} }
\end{figure}

In \cite{Calzone10} we summarized the current knowledge on the
interactions between cell fate decision mechanisms in a simplistic
wiring diagram (see Fig.~\ref{Model}) where a node represents
either a protein (TNF, FADD, FASL, TNFR, CASP8, cFLIP, BCL2, BAX,
IKK, NF$\kappa$B, CYT\_C, SMAC, XIAP, CASP3), a state of protein
(RIP1ub, RIP1K), a small molecule (ROS, ATP), a molecular complex
(Apoptosome, C2\_TNF, DISC\_FAS), a group of molecular entities
sharing the same function (BAX can thus represent either of BAX
and BAK, cIAP either cIAP1 or cIAP2, and BCL2 any of the BH1-4
BCL2 family members,…), a molecular process (Mitochondria
permeabilization transition, MPT, Mitochondrial outer membrane
permeabilization, MOMP) or a phenotype (Survival, Apoptosis,
Non-apoptotic cell death, NonACD). Each directed and signed edge
represents an influence of one molecular entity on another, either
positive (arrowed edge) or negative (headed edge).

The phenotype nodes on the diagram are simple interpretations of
the following molecular conditions: 1) activated NF$\kappa$B is
read as survival state; 2) lack of ATP is read as nonapoptotic
cell death state; 3) activated CASP3 is read as apoptotic cell
death. Absence of any of such conditions is interpreted as a
"naive`` cell state, corresponding to the fourth cellular
phenotype.

After extensive examination of the biological literature we
converted the diagram into a logical mathematical model of cell
fate decisions triggered by activation of cell death receptors.
The wiring diagram and the logical rules defining the model are
shown on Fig.~\ref{Model}.

By applying a technique adapted to discrete formalism
\cite{Naldi2009b}, we reduced this model to a $11$-dimensional
network, thus enabling a complete analysis of the asynchronous
dynamics (see \cite{Calzone10} for details).
This analysis identified 27 stable logical states and no
cyclic attractors.
Moreover, it showed that the
distribution of the stable logical states in the discrete
22-dimensional space of internal model variables (without
considering input and output variables) forms four compact
clusters, each corresponding to a particular cellular phenotype.
Three of these clusters can be attributed to a particular cell
fate (survival, apoptosis, necrosis) while the forth represents a
``naive'' survival state, where no death receptors are induced.

\section{Computing phenotype probabilities}

As we have already mentioned, the cellular fate decision machinery
is characterized by stochastic response, i.e. given a stimuli, the
cell can reach several final states, corresponding to different
phenotypes, with different probabilities. The role of mathematical
modeling in this case can be to predict these probabilities as
absolute values that can be matched to an experiment, or at least
to predict the relative changes of the probabilities after
introducing some perturbations to the system.

We have implemented this idea for the mathematical model of cell
fate decisions described above in the following manner.

In order to describe our results, let us introduce the notion of
asynchronous state transition graph.
On this graph, each node represents a state
of the system which in this case can be encoded by a $n$-dimensional
vector of 0s and 1s ($n$ being the dimension of the system).
A directed edge exists between two states $x$ and $y$ if there
exists an index $i\in\{1,\dots,n\}$ such that $y_i=f_i(x)\ne x_i$
and $y_j=x_j$ for $j\ne i$ (here, $f_i$ denotes the logical rule
of variable $x_i$, see Fig.~\ref{Model} for a complete list of the
model logical rules).
In principle, the state transition graph
could be defined independently
and without the biological diagram, however, this would require a
tremendous amount of empirical knowledge about the set of all
permissible transitions between the cell states which is not
available. Hence, the biological diagram with associated logical
rules is used as a compact representation and a tool to generate
the state transition graph. Detailed instructions on this
procedure can be found in \cite{Chaouiya2006,Tournier09}.

The set of all possible states provides a discrete phase space of
the system. The state transition graph contains all possible ways
of the systems dynamics (trajectories). In other words, it is the
{\it multidimensional epigenetic landscape} of the cell fate
decision system. Note that the state transition graph is assumed
to be rather sparse compared to the fully connected graph where
any two state transitions would be possible. Hence, on this
landscape, one can determine bifurcating states, points of no
return, etc.

The state transition graph allows to address the following
question: {\it Starting from a distinguished state of a cell, what
is the probability to arrive to each of the stable states?} In
biological terms: {\it Which proportions of a population of
resting cells exposed to death ligand will eventually display each
of the different phenotypes - cell fate}?

To answer the question, we converted the state transition graph
into a Markov process of random walk on a graph, following the
method described in \cite{Tournier09}.
To do that, we associated to each transition between two states a
probability (called transition probability). By applying classical
algorithms to the transition probability matrix (strongly connected
decomposition and topological sort), we obtained an {\it absorbing
discrete Markov chain}, and then analyzed it with classical
techniques \cite{Feller68}.

One of the critical points in such type of analysis lies in
the choice of the transition probabilities.
Once again, defining these probabilities
directly from some empirical observations is
impossible at present time. Hence, these probabilities should be
derived from the logical model with the use of some additional
assumptions.

The simplest assumption is to consider all transitions firing from
a given state as {\it equiprobable}. Biological interpretation of
such an assumption is not simple. In a way, we consider a
``generic'' cell in which all possible system trajectories take
place with equal probabilities (without dominance, i.e. any
preferable route). One can argue that in any particular concrete
cell, this would not be true anymore and that the generic cell is
not representative of anything real observed in any biological
experiment. Having in mind this difficulty, we avoid direct
interpretation of absolute values of probabilities, concentrating
rather on relative changes of them in response to some system
modifications such as removing a node or fixing a node's activity.
It happens that such a ``generic'' cell model is already capable of
reproducing a number of known experimental facts.

When the state transition graph is parametrized by transition
probabilities, one can
use standard techniques to compute the probability of hitting
a given stable state, considering that a random walk starts from a
given initial state. Then this probability is associated with a
probability of observing a particular phenotype in
given experimental conditions. For doing this, it is convenient to
define a unique initial state, which we choose to represent the
``physiological state'', the one representing un-induced cells
growing in a plate. In the model of Fig.~\ref{Model} it is the
state in which all elements are inactive except ATP, FADD and
cIAP. This is a stable state, which looses its stability when TNF
variable is changed from 0 to 1 and the dynamical system starts to
evolve in time.

\begin{figure}[htb]
\center{\includegraphics[width=80mm, angle=270]{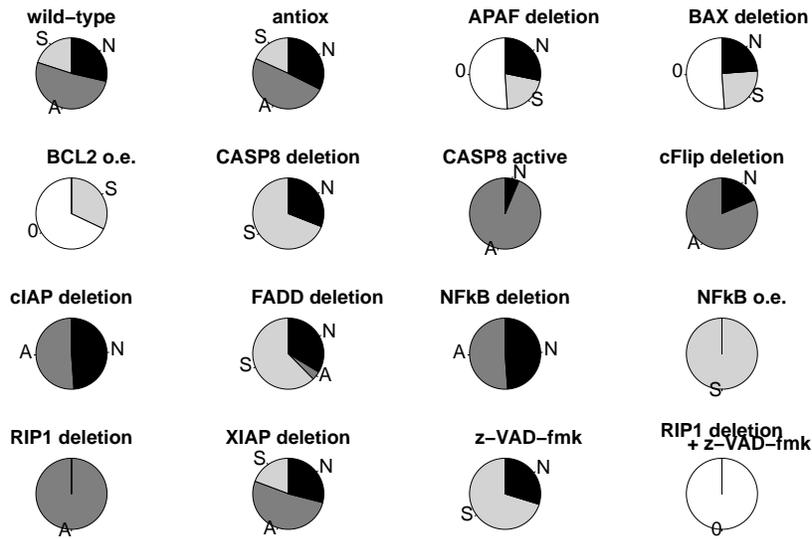}
\caption{\label{PieCharts} Changes in the phenotype probabilities
from the random walk on the state transition graph, starting from
the initial physiological state. Various ``mutant'' modifications
of the dynamical system are tested here. Here ``A`` denotes
Apoptosis, ``N`` denotes Necrosis and ``S`` denotes Survival,
``0'' denotes Naive state. ``O.e.'' stands for overexpression of a
protein, ``antiox'' corresponds to blunting the capacity of
NF$\kappa$B to prevent ROS formation, ``z-VAD fmk'' simulates the
effect of caspase inhibitor z-VAD-fmk. }}
\end{figure}

Using this approach, we performed a series of {\it in silico}
experiments in which the probability of arriving to stable states
was computed for the initial ("wild-type") model, or for a series
of modified (``mutant'') model. Typical model
modifications consisted in
fixing some nodes' activities to 0 or to 1. For our cell fate
decision model, the results are provided in Fig.~\ref{PieCharts}.
In \cite{Calzone10} this table was systematically compared with
the experimental data of the cell death phenotype modifications
observed in various mutant experimental systems, including cell
cultures and mice. The model was able to qualitatively
recapitulate all of them and to suggest some new yet unexplored
experimentally mutant phenotypes. The most interesting in this
setting would be to consider synthetic interactions between
individual mutants, when several nodes on the diagram are affected
by a mutation simultaneously.

\section{Identification of fragile points of the cell fate decision machinery}

Changing distribution of transition probabilities on the
asynchronous state transition graph can drastically change the
probabilistic outcome of a computational experiment. At the same
time, the probabilities for a random walk to converge to some
attractor
depend also on the structure of the state transition graph
which is determined solely from the discrete model. In order to
understand what are the critical determinants of a cellular
choice, we applied a novel strategy of discrete model analysis
consisting in parametrizing the state transition graph by changing
relative importance of certain
variables. In a certain sense, this strategy corresponds to
a sensitivity analysis, commonly applied for continuous models
based on ordinary differential equations and chemical kinetics
approach \cite{Turanyi1990}.

First of all, we postulate that our ``reference'' parametrization
corresponds to the equal probabilities of any possible transition
from a state. As mentioned earlier, this corresponds to a
``generic'' cell model,
where the relative speeds of all biochemical processes are assumed
equal. Mathematically, considering the dynamics as
a Markov process, all transitions from a given state $x$ to any of
its asynchronous successor are assigned equal probabilities
(if $x$ has $r$ successors, these probabilities are equal to
$1/r$). We will modify this default parametrization by
systematically changing relative speeds of certain
elements.
This will lead to some re-parametrization of the state transition
graph and consequent changes in the probabilities to reach
attractors.


The key idea of priority classes \cite{Faure2006,Naldi2009} consists in
grouping variables of a discrete model into classes according to
the speeds of the underlying processes
governing their turnover rates.
For instance, in the case of genetic regulatory networks, a natural
grouping consists in putting de novo protein synthesis
(transcription + translation) in a slow transition class in
comparison with other processes such as post-translational protein
modifications (phosphorylation, ubiquitination, ...) or complex
formation. Following this idea,
we can regroup nodes into priority classes to which some priority
ratios $w$ are assigned. Said differently, each variable
$x_i$ is assigned a priority value
$w_i$. For a given node, a value $w_i>1$ corresponds to a higher
than default priority, and a value $w_i<1$ to a lower than default
priority. The ratio $w_i$ can be interpreted as a global turnover
rate of the component represented by this node: those that are
produced (activated) and degraded (deactivated) fast will have a
large $w_i$.

Consider a state $x$, with $r$ asynchronous successors.
By definition, between
$x$ and each of its successors, one and only one
variable
can be updated. Let $y$ denote one of the successors of $x$, and
$i$ be the index of the corresponding updated variable. With the
uniform assumption described before, the probability of the
transition ($x\rightarrow y$) is independent of $i$ and is equal
to $1/r$. With priority classes, this probability is now weighted
by $w_i$, making the transition more probable if component $i$
belongs to a ``fast" class ($w_i$ greater than one) and less
probable if it belongs to a ``slow" class ($w_i$ less than one).
Obviously, for computing the actual transition probabilities
$p_{x\rightarrow y}$, a normalization should be applied so that:
$$\sum_{y \mbox{\scriptsize~succ. of }x}p_{x\rightarrow y} = 1.$$
Once the new values of the transition probabilities have been
computed, the same treatments as before can be applied, leading
to new values for the probabilities to reach the different
phenotypes, starting from a given initial condition.

\begin{figure}[htb]
\center{\includegraphics[width=60mm]{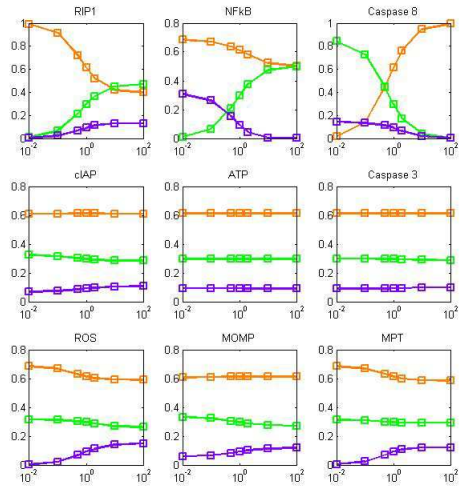}
\caption{\label{sensitivity_nodes}. Testing the effect of varying
node turnovers on the resulting phenotypic probabilities. The
absciss on the graphs shows the value of $w$ priority value, where
$w=1$ corresponds to the probabilities computed for the default
wild-type model (see Fig.~\ref{PieCharts}).
The colors are those adopted in \cite{Calzone10}: orange corresponds
to apoptosis, purple to necrosis and green to survival.}}
\end{figure}

This general method may be applied in two different ways. First,
one may use it to compute more realistic probabilities, that could
be compared to actual experimental results (the probability to
reach an attractor being compared with the proportion of cells
exhibiting the corresponding phenotype). However, such
calculations would need a complete classification of the relative
speeds of all biochemical mechanisms involved in the model. Given
the number and heterogeneity of these mechanisms, it is still
difficult to obtain such classification. Instead, we used the
method as a sensitivity analysis tool, in order to detect which
variables are more critical than others in the decision-making
process. Using the reduced model evoked earlier (see
\cite{Calzone10}), we considered each variable independently, and
successively boosted it or slowed it down by some multiplicative
factor. More precisely, to detect the sensitivity of the network
with respect to the turnover of variable $x_i$, we performed the
calculations for different values of $w_i$, the other weights
$w_j$ being kept at one (the reference value). By comparing the
probabilities to reach the three phenotypes -survival, apoptosis
and necrosis- with those of the initial model, one can detect
whether the system's response is sensitive or not to the turnover
rate of variable $x_i$. We performed such experiments for the nine
inner variables of the reduced model.
Figure~\ref{sensitivity_nodes} presents the results we obtained.

The plots
reveal several interesting properties. First, the most sensitive
components, which correspond to the curves with the highest
amplitude, are RIP1, NFkB and CASP8. This reinforces the idea that
these three components play a crucial role in the decision
process. This seems reasonable, especially for RIP1 and CASP8, as
they occupy an upstream position in the regulatory graph.
Interestingly, CASP3 turnover does not seem to be so important,
although CASP3 is a marker of apoptosis.
This confirms that even though CASP3 is essential for
the existence of apoptosis in the model (its removal completely
suppress apoptotic outcome, see Fig.~\ref{PieCharts}), its
turnover rate does not appear to be important in the dynamics of
the decision process (once it goes from 0 to 1, most of the
decision has already been made).
Remarkably, the turnovers of MOMP and MPT, both contributing to
the permeabilization of mitochondrial membrane, have different
effects: MOMP seems to affect mainly the decision between survival
and necrosis, while MPT plays a role in the switch between
apoptosis and necrosis.

The sensitivity analysis that is presented here is an extension of
the results proposed in \cite{Calzone10}. In contrast with the
all-or-none perturbations evoked in the previous part (where a
node is fixed to 0 or 1), here we consider finer perturbations by
modifying the turnover rates of the model's variables. A next step
would be to consider the relative strengths of the model's
interactions, instead of the model's variables. Such an approach
is currently investigated.

\section{Comparison with the fragilities exploited by cancer and its treatment}

Deregulations of the signalling pathways studied here can lead to
drastic and serious consequences. Hanahan and Weinberg proposed
that escape of apoptosis, together with other alterations of
cellular physiology, represents a necessary event in cancer
promotion and progression \cite{Weinberg2011}. As a result,
somatic mutations leading to impaired apoptosis are expected to be
associated with cancer. In the cell fate model presented here,
most nodes can be classified as pro-apoptotic or anti-apoptotic
according to the results of ``mutant'' model simulations, which are
correlated with experimental results found in the literature.
Genes classified as pro-apoptotic in our model include caspases-8
and -3, APAF1 as part of the apoptosome complex, cytochrome c
(Cyt\_c), BAX, and SMAC. Anti-apoptotic genes encompass BCL2,
cIAP1/2, XIAP, cFLIP, and different genes involved in the NFkB
pathway, including NFKB1, RELA, IKBKG and IKBKB (not explicit in
the model). Genetic alterations leading to loss of activity of
pro-apoptotic genes or to increased activity of anti-apoptotic
genes have been associated with various cancers. Thus, we can
cross-list the alterations of these genes deduced from the model
with what is reported in the literature and verify their role and
implications in cancer.

\begin{figure}
\center{\includegraphics[width=120mm]{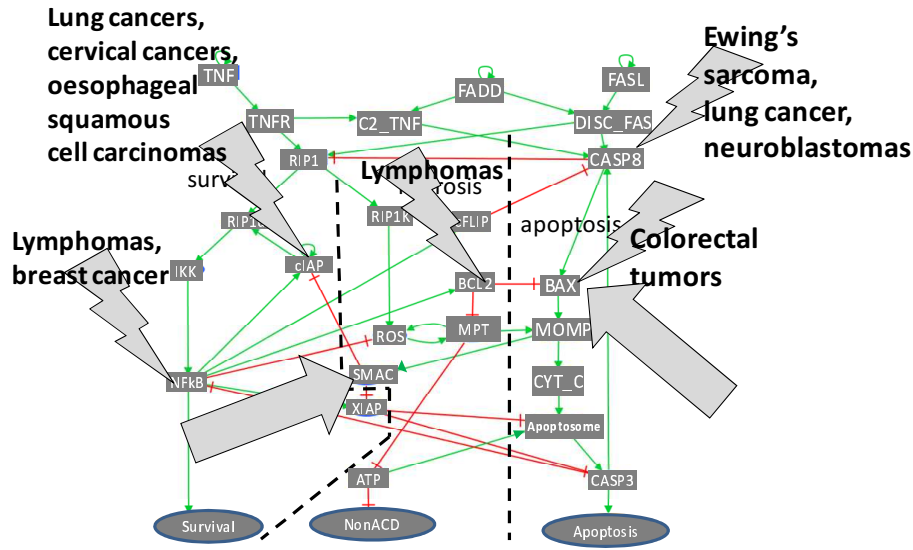}
\caption{\label{CancerHits} Cell fate decision fragilities
identified in various cancers. Flash arrows, hitting from left to
right, represent overexpression or amplification, those hitting
from right to left show deletion and down-regulation). Rectangular
arrows point to components targeted by cancer treatment strategies
(SMAC and BAX mimetics). }}
\end{figure}

For instance, concerning pro-apoptotic genes, frameshift mutations
in the ORF of the BAX gene are reported in $>50$\% of colorectal
tumours of the micro-satellite mutator phenotype \cite{Rampino97}.
Expression of CASP8 is reduced in $\approx$24\% of tumours from
patients with Ewing's sarcoma \cite{Lissat07}. Caspase-8 was
suggested in several studies to function as a tumour suppressor in
neuroblastomas \cite{Teitz01} and in lung cancer
\cite{Shivapurkar02} (see Fig.~\ref{CancerHits}).

On the other hand, constitutive activation of anti-apoptotic genes
is often observed in cancer cells. The most striking example is
the over-expression of the BCL2 oncogene in almost all follicular
lymphomas, which can result from a t(14;18) translocation that
positions BCL2 in close proximity to enhancer elements of the
immunoglobulin heavy-chain locus \cite{Croce08}. As for the
survival pathway, elevated NFkB activity, resulting from different
genetic alterations or expression of the v-rel viral NFkB isoform,
is detected in multiple cancers, including lymphomas and breast
cancers \cite{Karin02}. An amplification of the genomic region
11q22 that spans over the cIAP1 and cIAP2 genes is associated with
lung cancers \cite{Dai03}, cervical cancer resistance to
radiotherapy \cite{Imoto02}, and oesophageal squamous cell
carcinomas \cite{Imoto01} (see Fig.~\ref{CancerHits}).

Some of the components of the cell fate decision machinery are
considered currently for the use in cancer treatment in
pre-clinical or clinical trials. To give some examples, SMAC
mimetics directly target dysregulated, neoplastic cells that
overexpress IAPs or underexpress SMAC \cite{Chen09}. BCL-2
inhibitors, most notably BAX mimetics, are currently passing
clinical trials (for example, see \cite{Ready11}).

In our sensitivity analysis, the variables NF$\kappa$B and CASP8
appear among the most "vulnerable" components of the cell fate
decision machinery, which could explain why the gene products they
represent are fragile points used by cancer.

BLC-2 does not show up as a much sensitive node in the model.
However, its direct target, MPT is a fragile site, accordingly to
our analysis. Also analysis of our model shows that RIP1 is a
powerfull and sensitive switch able to reverse phenotype
probabilities. Until so far we are not aware about possible
targeting of RIP1 functions in cancer treatment, which can be
explained by still relatively poor characterization of its
substrates and difficulties connected with targeting specific RIP1
activities.

\section{Conclusion}

Mathematical models provide a way to test biological hypotheses
{\it in silico}. They recapitulate consistent heterogeneous
published results and assemble disseminated information into a
coherent picture using an appropriate mathematical formalism
(discrete, continuous, stochastic, hybrid, etc.), depending on the
questions and the available data. Then, modeling consists of
constantly challenging the obtained model with available published
data or experimental results (mutants or drug treatments, in our
case). After several refinement rounds, a model becomes
particularly useful when it can provide counter-intuitive insights
or suggest novel promising experiments.

Here, we have conceived a mathematical model of cell fate
decision, based on a logical formalization of well-characterized
molecular interactions. Former mathematical models only considered
two cellular fates, apoptosis and cell survival \cite{Lavrik2010}.
In contrast, we include a non-apoptotic modality of cell death,
mainly necrosis, involving RIP1, ROS and mitochondria functions.

By analyzing properties of the state asynchronous transition
graphs associated with the discrete model, we implemented a
procedure to simulate the process of stochastic cellular decision
making in response to activation of death receptors. These
simulations were able to predict relative changes for
probabilities of cellular phenotypes in response to some system
perturbations such as a knock-out of a gene or treatment with a
drug. These predictions happened to be fully compatible with
published data from mouse experiments, and provided new
predictions to be tested.

Moreover, on this model we have tested a novel strategy of
discrete model analysis, consisting in finding fragile or most
sensitive places of the cell fate decision machinery.
Changing the cellular parameters determining choices made at these
fragile sites affect the probabilities for a cell to reach a
particular cellular phenotype. We found out that this type of
analysis can explain some of the common fragilities associated
with tumorigenesis and also with currently employed cancer
treatment strategies.

\begin{acknowledgement}
We would like to acknowledge support by the APO-SYS EU FP7
project.
A. Zinovyev, S. Fourquet, L. Calzone and E. Barillot
are members of the team ``Systems Biology
of Cancer'', Equipe labellisee par la Ligue Nationale Contre le
Cancer.
L. Tournier is member of the Systems Biology team in the laboratory
MIG of INRA (french institute for agronomical research).
The study was also funded by the Projet Incitatif
Collaboratif ``Bioinformatics and Biostatistics of Cancer'' at
Institut Curie.
\end{acknowledgement}

%
%
%

\end{document}